\documentclass[conference]{IEEEtran}
\IEEEoverridecommandlockouts

\usepackage{cite}
\usepackage{amsmath,amssymb,amsfonts}
\usepackage{algorithmic}
\usepackage{graphicx}
\usepackage{textcomp}
\usepackage{xcolor}

\usepackage{braket}
\usepackage{hyperref}

\newcommand{\noeditingmarks}{}

\ifx\noeditingmarks\undefined
    \newcommand{\blue}[1]{\textcolor{blue}{#1}}
\else
    \newcommand{\blue}[1]{#1}
\fi

\ifx\noeditingmarks\undefined
    
\else
    
\fi

\newcommand{\para}[1]{\smallskip \noindent\textbf{#1}}
\newcommand{\softpara}[1]{\smallskip \noindent \underline{#1}}

\newcommand{\yb}{\texttt{Yb}}

\newcommand{\uw}{\texttt{$\mu$W}}

\newcommand{\mpfont}{\scriptsize}

\ifx\noeditingmarks\undefined
    \newcommand{\MPworker}[2]{{\color{#1}\vrule\vrule}{\marginpar{\color{#1}\mpfont #2}}}
\else
    \newcommand{\MPworker}[2]{}
\fi

\begin{document}

\bstctlcite{IEEEexample:BSTcontrol}

\title{Simulation of a Heterogeneous Quantum Network}

\author{


\IEEEauthorblockN{Hayden Miller$^*$, Caitao Zhan$^{\dagger}$, Michael Bishof$^{\dagger}$, Joaquin Chung$^{\dagger}$, Xu Han$^{\dagger}$, Prem Kumar$^{\ddagger}$, Rajkumar Kettimuthu$^{\dagger}$}

\IEEEauthorblockA{$^{*}$Brown University (USA), $^{\dagger}$Argonne National Laboratory (USA), $^{\ddagger}$Northwestern University (USA)}
    
\thanks{This material is based upon work supported by the U.S. Department of Energy, Office of Science, Advanced Scientific Computing Research (ASCR) program under contract number DE-AC02-06CH11357 as part of the InterQnet quantum networking project. }   

}

\maketitle

\begin{abstract}
    Quantum networks are expected to be \emph{heterogeneous} systems, combining distinct qubit platforms, photon wavelengths, and device timescales to achieve scalable, multiuser connectivity. 
    Building and iterating on such systems is costly and slow, which motivates hardware-faithful simulations that explore architecture design space and justify implementation decisions. 
    This paper presents a framework for simulating heterogeneous quantum networks based on SeQUeNCe, a discrete-event simulator of quantum networks. 
    We introduce faithful device models for two representative platforms--Ytterbium atoms and superconducting qubits-- to implement entanglement generation and swapping protocols for time-bin encoded photons.
    Using extensive simulations that account for disparate clock rates and quantum frequency conversion and transduction losses/noise, we map the rate–fidelity trade space and identify the dominant bottlenecks unique to heterogeneous systems. 
    The models are open source and extensible, enabling reproducible evaluation of future heterogeneous designs and protocols. 
\end{abstract}


\section{Introduction}
\label{sec:introduction}

Quantum communication holds enormous potential for many ground-breaking scientific and technological advances. 
A quantum network~\cite{wehner2018quantum} serves as the backbone for applications such as the transmission of an unknown quantum state~\cite{hu-teleportation-nrp23}, quantum key distribution~\cite{bennett-qkd-14},
distributed quantum computing~\cite{calaffi-dqc-cn24}, and distributed quantum sensing~\cite{hillery-qsn-pra23}. 
Researchers have demonstrated early prototypes of campus-scale~\cite{oakridge-prx-21,interqnet-25,quantnet-24} and metropolitan-scale~\cite{longisland-24,bersin_development_2024,delft_testbed-science-24} quantum communication networks.
Despite the recent advances in this space, achieving the scalability needed for efficient communication encompassing a large number of nodes and users remains a challenge. 

A scalable and useful quantum network is expected to be \emph{heterogeneous}, involving a diversity of quantum devices, which would need a robust control layer and protocols to maintain sufficiently high fidelities and entangled pair rate.
Various quantum technologies are being developed in parallel and they are expected to co-exist because each one provides unique advantages. 
The majority of quantum network testbeds are composed of homogeneous quantum technologies.
Researchers in~\cite{quantnet-24} use trapped ions as the qubit platform to build a quantum network, while researchers in~\cite{delft_testbed-science-24} use diamond spin qubits.
In recent years, developing a heterogeneous quantum network has gained interest due to the belief that heterogeneity is the pathway towards a scalable quantum internet.
For example, the InterQnet project is developing a first-of-a-kind heterogeneous quantum network testbed~\cite{interqnet-25}, comprised of neutral atom, atom-like defect, and single-electron qubits.

Developing a quantum network is a challenging endeavor, and heterogeneity introduces an additional layer of complexity.
In a homogeneous quantum network, all nodes and links use the same qubit platform, wavelengths, and protocols--so one can design a single stack and replicate it. 
At the physical and link layers, a heterogeneous quantum network mixes platforms (e.g., NV centers, trapped ions, superconductors, atoms), memories, and photon wavelengths.
At the network and control planes, a heterogeneous quantum network introduces clock and synchronization challenges because each qubit platform works at different timing regimes.

Because heterogeneous quantum network testbeds are costly, slow to iterate, and hard to instrument, simulation studies with hardware-faithful models are the only practical path to explore the architecture design space, justify implementation decisions, standardize protocols, and predict network performance with futuristic parameters.
Motivated by the above, this paper aims to simulate a heterogeneous quantum network using SeQUeNCe~\cite{sequence}, an open-source and customizable quantum network simulator.
We built new modules in SeQUeNCe that accurately model devices and network protocols to generate remote entanglement in a heterogeneous quantum network.
Through extensive simulations, we discovered some interesting insights. First, using current experimental parameters~\cite{covey}, we found the optimal attempts per reload for an \yb-\yb\ link is $\sim65$. Second, we found that compared to the \yb-\yb\ link, the \yb-\uw\ had a higher entanglement generation rate but a lower entanglement fidelity. Finally, we showed that for larger \yb\ and \uw\ networks, the \uw\ coherence time proved to be the biggest bottleneck in remote entanglement generation.
This project is open-source at~\cite{github}.

\begin{figure*}[t]
    \centering
    \includegraphics[width=0.8\linewidth]{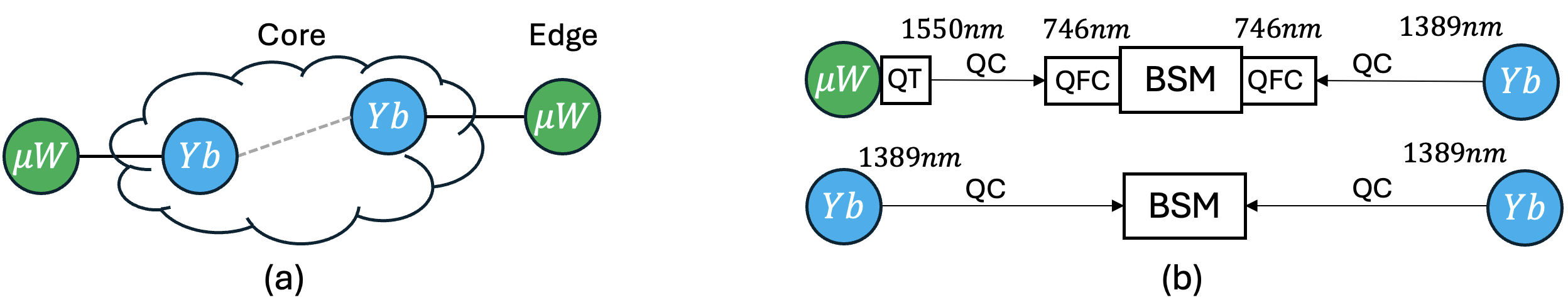}
    \vspace{-0.15in}
    \caption{(a) Heterogeneous quantum network. \yb-nodes serve as quantum repeaters at the network core, while \uw-nodes serve as quantum computing nodes at the network edge. (b) Hardware components of both heterogeneous and homogeneous links, including quantum channels (QC), quantum frequency converters (QFC), time-bin Bell State Measurement (BSM) devices, quantum transducers (QT), and classical channels (not shown in the figure).}
    \vspace{-0.1in}
    \label{fig:heterogeneous}
\end{figure*}

\para{Contributions and Paper Organization}. We introduce the problem of simulating a heterogeneous quantum network and discuss the related work in~\S\ref{sec:problem}. Our key contributions are:
\begin{enumerate}
    \item We introduce four device models in~\S\ref{sec:device}: Ytterbium memory, microwave memory, quantum frequency converter, and quantum transducer -- building blocks of a heterogeneous quantum network.
    \item We present the SeQUeNCe implementation of device models and network protocols that coordinate heterogeneous remote entanglement generation in~\S\ref{sec:protocol}.
    \item We conduct extensive simulations of the heterogeneous quantum networks and discuss our findings in~\S\ref{sec:result}.
\end{enumerate}

\section{Problem and Related Work}
\label{sec:problem}

In this section, we introduce the problem of simulating a heterogeneous quantum network and discuss the related work.

\subsection{Problem}
We aim to solve the problem of modeling devices and developing network protocols that serve as the building blocks of a heterogeneous quantum network. 
The developed device models and network protocols are incorporated into a discrete event quantum network simulator called SeQUeNCe, which we use to study how the hardware device parameters affect the performance of the heterogeneous quantum network by conducting comprehensive simulations. 

Figure~\ref{fig:heterogeneous} illustrates our heterogeneous quantum network simulation that consist of two types of nodes containing different qubit technologies. \yb\ nodes are based on neutral Ytterbium atoms trapped in optical tweezer arrays~\cite{covey}, while \uw\ nodes are based on transmon qubits~\cite{transmon}. 
Heterogeneous nodes are connected via telecom fiber, and they emit time-bin encoded photons (at different frequencies) to generate entanglement using the meet-in-the-middle (MIM) protocol.
Thus, quantum frequency converters are needed to convert photons of different frequencies to a common frequency before interfering at the Bell state measurement (BSM) node.
Quantum transducers are needed on the \uw\ nodes to convert microwave photons emitted by the transmons to the optical wavelengths necessary for telecom fiber.


\para{Challenges.}
Simulating a heterogeneous quantum network poses additional challenges beyond those of homogeneous settings. 
First, devices such as quantum memories differ in physical implementation (e.g., \yb, \uw) and operate at distinct frequencies, coherence times, and interface efficiencies. 
Quantum frequency converters and quantum transducers add an extra layer of complexity to the simulation.
Accurately modeling such devices requires abstract yet realistic representations of their quantum dynamics and noise behaviors.
For example, quantum frequency converters will produce noise photons while converting the frequencies of the signal photon. And how the noise photons affect the entanglement pair is worth investigating.
Second, network protocols (e.g., entanglement generation) must adapt dynamically to device heterogeneity (i.e., photon emission frequency), creating a coupling between hardware modeling and protocol logic. 


\subsection{Related Work}

\subsubsection{Heterogeneous quantum networks}
Most quantum network testbeds are homogeneous, where all nodes share identical qubit platforms and interface standards (e.g., trapped ions~\cite{quantnet-24}, neutral atom ensemble~\cite{longisland-24}, and diamond spin~\cite{delft_testbed-science-24}).
Similarly, most theoretical studies on quantum networks assume homogeneous network nodes for simplicity~\cite{swappingtree-tqe-22,acp-qcnc-25,graphstate-tqe-25}. 
However, recent research has caught interest in heterogeneous quantum networks, which integrate diverse physical platforms—such as trapped ions, superconducting qubits, nitrogen-vacancy centers, atomic ensembles, and photonic systems to leverage their complementary advantages~\cite{heterogeneous_routing-25,interqnet-25}. 
For example, Pandey et al.~\cite{hetero-qce-25} studied entanglement routing in a heterogeneous network, where heterogeneity is reflected in functionality (general-purpose quantum computer vs. specialized quantum router) and quality of service (probability of success).
\blue{Yet, networks exhibiting heterogeneity in qubit implementation have largely been unexplored.}

\subsubsection{Device modeling in discrete event quantum network simulators}
Discrete event quantum network simulators~\cite{sequence,netsquid,quisp} are important tools to study quantum networks, including quantum network protocols~\cite{quantum-conext-20}, quantum network architectures~\cite{qlan-icc-25}, as well as quantum hardware devices~\cite{absorptive-qce-22}.
In~\cite{transducer-qce-25}, quantum transducers are modeled as an interface between superconducting nodes across an optical network.
In~\cite{atom-qce-23}, atomic ensemble quantum memories are modeled to study how the BDCZ protocol entanglement generation rate and fidelity depend on atomic ensemble parameters.
\begin{figure*}[ht]
    \centering
    \includegraphics[width=0.9\textwidth]{figure/yb.png}
    \vspace{-0.15in}
    \caption{Ytterbium (\yb) Node. (a) shows energy levels, atomic states, and atom transitions during the initialization, preparation, and generation steps\cite{covey}. Each energy level corresponds to a manifold in Yb specified by term symbols (e.g. $^3D_1$) and nuclear spin (e.g. $F=\frac{3}{2}$). A time-bin encoded photon at $1389~nm$ is emitted and collected by the fiber. 
    (b) shows the steps of the time-bin entanglement generation and their timing. In particular, the preparation and generation steps are illustrated in further detail. 
    (c) explains the symbols used in (b) and includes the transition details in (a).}
    \vspace{-0.1in}
    \label{fig:ytterbium}
\end{figure*}

\section{Physical Layer Device Models}
\label{sec:device}

In this section, we introduce our device models that serve as the building blocks of a heterogeneous quantum network.

\subsection{Ytterbium (\yb) Atom Memory}

Our \yb\ memory is composed of $^{171}$\yb\ atoms where each qubit is represented by the nuclear spin of an atom $(\ket{\downarrow}, \ket{\uparrow})$. \yb\ atoms are of particular interest because they have metastable ``clock" states with a 20-second lifetime and have a transition that emits 1389~nm wavelength photons in the telecom E-Band. 
We model our \yb\ device based on work by Li et al.~\cite{covey}.
To model time-bin entanglement generation, we extract a series of timing and efficiency parameters of device control operations, and track the atomic state of each qubit.
The atomic state keeps track of whether the atom is still trapped in the node (trapped atoms can be lost and re-trapping requires time), and the quantum state of the atom. 
Tracking timing, efficiency, and qubit states is essential towards understanding entanglement rate and fidelity, and thus the viability of heterogeneous quantum networks for a given set of device parameters. 
We identify a total number of \textbf{24 parameters} for our \yb\ model, spread across entanglement generation, swapping, and readout operations.

\subsubsection{\yb-photon Entanglement Generation}
There are five steps during the time-bin \yb-photon entanglement generation, illustrated in Figure~\ref{fig:ytterbium}(b) and explained below.

\begin{figure*}[h]
    \centering
    \includegraphics[width=0.9\linewidth]{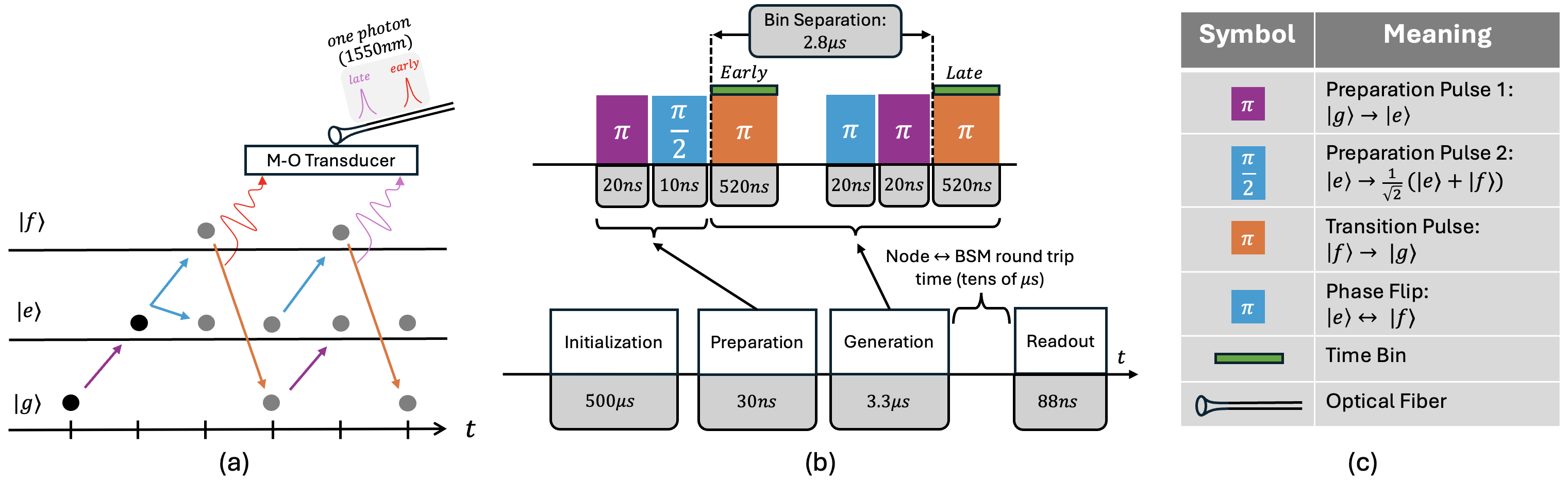}
    \vspace{-0.15in}
    \caption{Microwave (\uw) Node. (a) shows the energy levels and transitions of the transmon during initialization, preparation, and generation. After the microwave-optical (M-O) transduction, a time-bin encoded photon at $1550~nm$ is sent to the fiber. (b) shows the steps of the time-bin entanglement generation and their timing. The preparation and generation steps are illustrated in further detail. (c) explains the symbols used in (b) and the transition details in (a).}
    \vspace{-0.1in}
    \label{fig:superconducting}
\end{figure*}

\softpara{Reset}. The Reset stage creates an array of Yb atoms and takes $500~ms$ to complete. This process starts with the use of an oven to create a dilute atomic vapor of Yb atoms inside the room-temperature vacuum system of the device\cite{neutralatom}. Atoms are then laser cooled, trapped in optical tweezers, and rearranged to form a defect-free array of qubits. 
This must occur at initialization of the device, but through various processes, atoms eventually fall out of the tweezers. 
In~\cite{covey}, the reset stage is repeated every 128 entanglement generation attempts to reload the register.
The dotted border of the ``Reset'' box in Figure \ref{fig:ytterbium}(b) indicates this occurs periodically. 

\softpara{Initialization}. During the Initialization phase, we set a given Yb atom to the ground state $\ket{g}$. To overcome the 20 second lifetime of the ``clock'' states ($\ket{e_\downarrow},\ket{e_\uparrow}$), the device undertakes a depumping procedure, moving any probability the atom is in $\ket{e_\downarrow}$ or $\ket{e_\uparrow}$, to $\ket{g}$.
Depumping takes $51.4~\mu s$ to complete, with a $3\%$ probability the atom decays to a non-trappable state and is lost.


\softpara{Cooling}. To keep the atom temperature below the tweezer trap's energy depth, a $1.4~ms$ sub-doppler cooling (gray molasses and then electromagnetically induced transparency) is required. 

\softpara{Preparation} begins with a clock $\pi$ pulse elevating the atom from $\ket{g}$ to $\ket{e_{\uparrow}}$, one of our qubit states. This is followed by a Raman $\frac{\pi}{2}$ pulse sending $\ket{e_{\uparrow}}\rightarrow \frac{1}{\sqrt2}(\ket{e_{\uparrow}} + \ket{e_{\downarrow}})$, our desired starting \emph{equal superposition state}. 
The parameters here are the required time ($5.3~\mu s$) and the final state of our atom.

\softpara{Generation} is the phase yielding atom-photon entanglement. 
It begins with the early telecom excitation $\pi$ pulse that drives $\ket{e_{\downarrow}}\rightarrow \ket{f}$ while leaving $\ket{e_{\uparrow}}$ unchanged. 
If the atom is promoted to $\ket{f}$, it decays back to $\ket{e_{\downarrow}}$ with a probability of $64\%$, emitting a $1389~nm$ telecom photon. 
Alternatively, the atom decays to $\ket{p}$ and then to $\ket{g}$ with probability of $35\%$, emitting a non-1389~nm photon and remaining trapped. With probability of $1\%$, the atom decays to the non-trappable $^3P_2$ and is lost. 
The decay from $\ket{f}$ to any of the three options/branches is not instantaneous, since the $\ket{f}$ state has a lifetime of $330~ns$. To ensure a high probability of emission, the time-bin is given a width of $520~ns$~\cite{covey}. 
This is the time window over which we look for a click at the BSM node detectors. 
However, even for a correct decay path, there remains a $21\%\hspace{1mm}(e^{-\frac{520}{330}})$ chance that no emission occurs during the time-bin.
Our model tracks the branching ratios, $\ket{f}$ lifetime, and time-bin width as key parameters in our model.

After a $2.1~\mu s$ delay to nearly guarantee decay from $\ket{f}$~\cite{covey}, a Raman $\pi$ pulse ($0.7~\mu s$) is applied, phase flipping the qubit $\ket{e_{\downarrow}}\leftrightarrow \ket{e_{\uparrow}}$. This cumulative time of $2.8~\mu s$ is referred to as the `bin separation', as it is the time between successive excitation pulses that begin each time-bin. 
The phase flip lets $\ket{e_\uparrow}$ --the portion of the atomic state that did not emit during the early attempt-- become $\ket{e_\downarrow}$, which can be excited in the late window.
Then, a second (late) telecom $\pi$ pulse is applied, again driving $\ket{e_{\downarrow}}\rightarrow \ket{f}$.
This allows a telecom photon to be emitted in the late time-bin, but only if no photon was emitted earlier. 
Because the early and late excitations act on different components of the initial atomic state, the atom emits only one photon during the entire process.
This series of operations yields time-bin atom-photon entanglement in the joint state of $\frac{1}{\sqrt 2}(\ket{e_{\uparrow}E} + \ket{e_{\downarrow}L})$, where $\ket{E}$ and $\ket{L}$ correspond to a photon emission in the early and late time-bins. Here, we track the bin separation time and each of the pulse times as key parameters.


\subsubsection{Swapping}
For a \yb\ repeater, entanglement swapping simply involves a Bell state measurement of neighboring atoms within the repeater. We model this with the swapping fidelity (0.98) parameter.


\subsubsection{Readout} 
The readout is partial quantum state tomography using an $X$ or $Z$ basis measurement by means of fluorescent imaging taking $37.5~ms$ and with fidelity of $0.995$. 


\subsection{Microwave (\uw) Memory}
\label{mW node}

Our microwave memory is a model of a transmon coupled to coplanar waveguide resonators and an on-chip quantum transducer, enabling communication over optical wavelengths. The transmon qubit is implemented by its first two energy levels ($\ket{g}$ and $\ket{e})$, while its third ($\ket{f}$) enables generation of a microwave photon entangled to the transmon. 
Despite speed and gate fidelity advantages of transmons, 
their microwave-frequency emissions need to be converted to fiber-viable optical wavelength using a quantum transducer -- a currently inefficient and noisy technology.
We primarily model our \uw\ device based on the work in \cite{transmon} and \cite{transducer-review} on transmon-photon entanglement and quantum transduction. For our \uw\ model, we identified a total number of \textbf{14 parameters} spread across entanglement generation, swapping, and readout.



\subsubsection{\uw-photon Entanglement Generation}
The time-bin \uw-photon entanglement generation process requires three steps, two steps fewer than the \yb\ node (see Figure~\ref{fig:superconducting}(b)) as reset and cooling phases are unnecessary.

\softpara{Initialization.} To begin the entanglement generation process on a \uw\ node, the qubit must be initialized in $\ket{g}$. For this step we only have an initialization time parameter ($500~\mu s$).


\softpara{Preparation.} To prepare our transmon for qubit-photon entanglement, a $20~ns$ $\pi$-pulse is applied~\cite{transmon}, sending $\ket{g} \rightarrow \ket{e}$, followed by a $10~ns$ $\frac{\pi}{2}$-pulse, sending $\ket{e} \rightarrow \frac{1}{\sqrt2}(\ket{e} + \ket{f})$.

\softpara{Generation.} The qubit-photon entanglement generation process on a \uw\ node can take hundreds of nanoseconds. Yet, we are interested in simulating $Yb-\mu W$ links, with heralded entanglement via optical BSM devices.
To achieve photon interference, we need the time-bin width and bin separation parameters of heterogeneous node to match. As the \yb\ node is the restricting device here, we inherit its time-bin width of $520~ns$ and bin separation of $2.8~\mu s$. 

To generate microwave photons, a Raman-type $\pi$ pulse is applied transitioning $\ket{f} \rightarrow \ket{g}$ and adding a photon to the resonator if the state was in fact in $\ket{f}$. This pulse can be done in only $200~ns$, yet we choose to shape it to last $520~ns$ for a proper time-bin width. 
To enable a second ``late'' emission pulse, the portion of the transmon state that was in $\ket{e}$ before the transition is now elevated to $\ket{f}$ through a $20~ns$ $\pi$-pulse. 
Next, the portion of the transmon that transitioned down to $\ket{g}$ is excited up to $\ket{e}$ through the same $20~ns$ $\pi$-pulse used during preparation. 
Now, the transmon is once again in $\frac{1}{\sqrt2}(\ket{e} + \ket{g})$ but with the probabilities reversed. 
Then, the Raman-type pulse is repeated, transitioning $\ket{f} \rightarrow \ket{g}$, and adding a photon to the resonator if the state was in fact in $\ket{f}$ (which earlier was $\ket{e}$, the non-emitting state). 
This yields transmon-photon entanglement in the joint state of $\frac{1}{\sqrt 2}(\ket{eE} + \ket{gL})$ where $E/L$ denote photons emitted during the early/late time-bins, respectively.
To achieve the bin separation time of $2.8~\mu s$, our \uw\ model waits for the last $40~ns$ of this window to begin the two consecutive $\pi$ pulses. Due to a $T_1$ coherence time of $500~\mu s$, the portion of the quantum state in $\ket{e}$ decoheres to $\ket{g}$ with a probability of $0.0056$. Decoherence, collapses the transmon state to $\ket{g}$ and prevents a possible late photon. 

Once in the resonator, an emitted photon is converted from microwave to optical frequency through an on-chip transducer. The transducer is a one stage piezo-optomechanical platform with electrical and mechanical modes both operating at $10~GHz$. The transducer has two key parameters, efficiency and added noise. We chose a default efficiency of $60\%$ to enable reasonable fidelities and rates, though we acknowledge this is an ambitious number. 
\blue{We set the mean number of photons added by thermal noise during conversion (across each time bin) to $0.047$ (see equations 9-10 in \cite{transducer-review}).}



\subsubsection{Swapping}
As \uw\ nodes will only serve as edge nodes in our networks, we only need to model the correction process during entanglement swapping (possible $X$ gate). Thus, the only key parameter here is the $\ket{g} \leftrightarrow \ket{e}$ $\pi$ pulse time ($20~ns$).

\subsubsection{Readout}
The readout is partial quantum state tomography using an $X$ or $Z$ basis measurement taking $88~ns$ and with fidelity of $0.992$ \cite{uw_readout}.

\subsection{Quantum Frequency Converter (QFC)}

\begin{figure}
    \centering
    \includegraphics[width=0.9\linewidth]{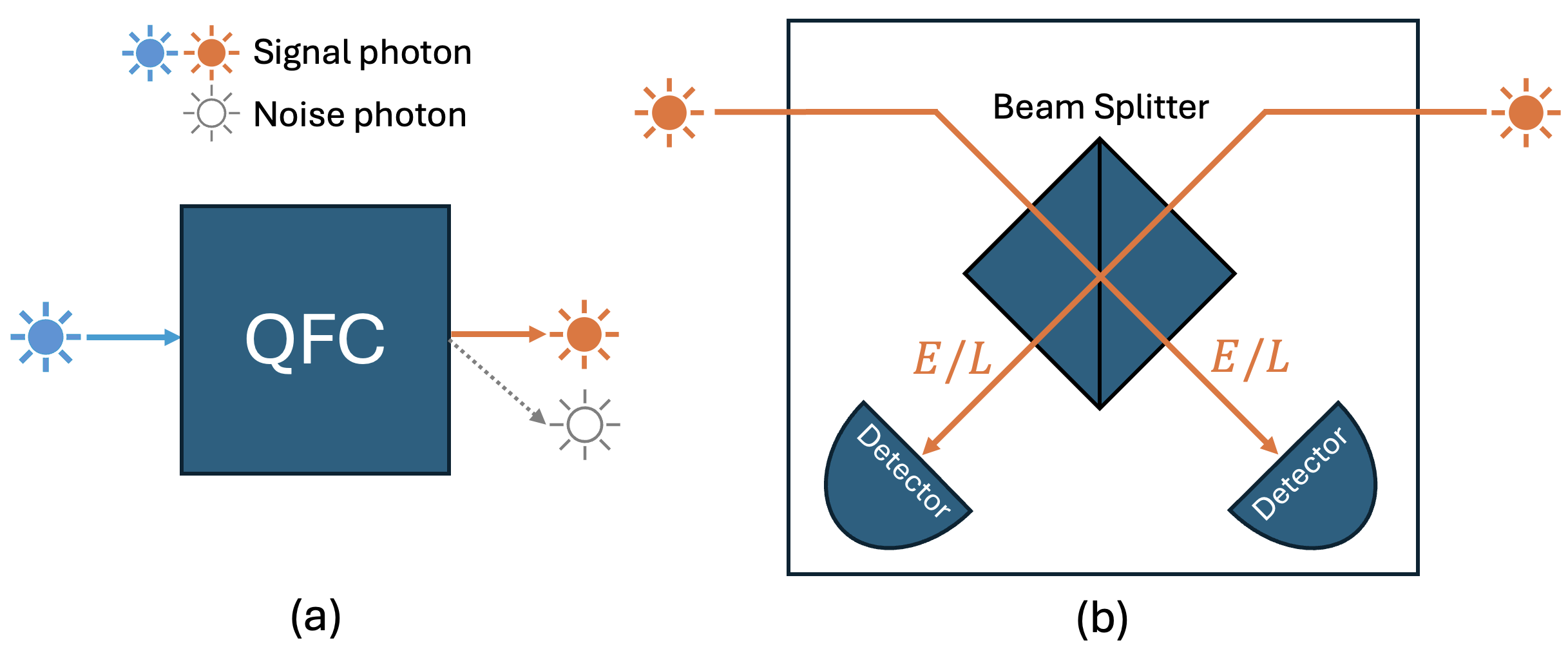}
    \vspace{-0.15in}
    \caption{(a) Quantum frequency converter. Signal photons in different colors reflect different frequencies, while gray non-filled photons are noise photons from the QFC. (b) Time-bin BSM node}
    \vspace{-0.1in}
    \label{fig:qfc-bsm}
\end{figure}

A QFC converts an input signal photon from one frequency to another, potentially failing and/or generating a noise photon (see Figure~\ref{fig:qfc-bsm}(a)).
Different hardware modalities emit photons at different wavelengths, whereas a BSM node requires two input photons that are the same in frequency.
For example, \yb\ atoms use the ${}^{3}D_1\rightarrow {}^{3}P_0$ telecom-band optical transition which emits photons at $1389~nm$, while most quantum transducers designed for superconducting qubits convert microwave excitations into $1550~nm$ photons in the telecom C-band. 
Therefore, QFCs are essential in heterogeneous quantum networks for bridging disparate wavelengths and enabling high-visibility interference at the BSM node.
In Figure~\ref{fig:heterogeneous}(b), two QFCs convert $1389~nm$ and $1550~nm$ to $746~nm$ at the BSM node, because converting from telecom to visible light is easier than converting between $1389~nm$ and $1550~nm$.

A QFC has \textbf{4 parameters}: conversion success probability ($99\%$), noise photon rate ($0.5\%$), input photon frequency, and output photon frequency.
The conversion success probability is the probability of successfully converting a photon, and the noise photon ratio is the probability of generating an unwanted noise photon \blue{along with the signal photon on each attempt.} 


\subsection{Time-bin Bell-state Measurement (BSM) Node}

A time-bin BSM node interferes two incoming time-bin-encoded photons from remote quantum nodes and projects their associated matter qubits into an entangled state.
\blue{This} requires a 50:50 beam splitter, an unbalanced Mach-Zehnder interferometer (shown as $E/L$ in Figure~\ref{fig:qfc-bsm}(b)) that overlaps the early and late components, and time-resolved single photon detectors.
In the ideal case (without noise photons and detector dark counts) \blue{one click in the early time-bin and one in the late time-bin heralds the $\psi^+$ state between the matter qubits.}


A BSM node has \textbf{3 parameters}: detector efficiency ($85\%$), detector detection rate ($25*10^6 ~Hz$), and dark count rate ($11$ Hz). Detector efficiency is the probability a photon is successfully detected; detection rate is the maximum photon detection rate, one over detection rate is the detection window ($40ns$); and the dark count rate is the frequency a detector incurs a false detection when no actual photons are present. 



\section{SeQUeNCe Extension and Customization}
\label{sec:protocol}

\begin{figure}[th]
    \vspace{-0.1in}
    \centering
    \includegraphics[width=0.7\linewidth]{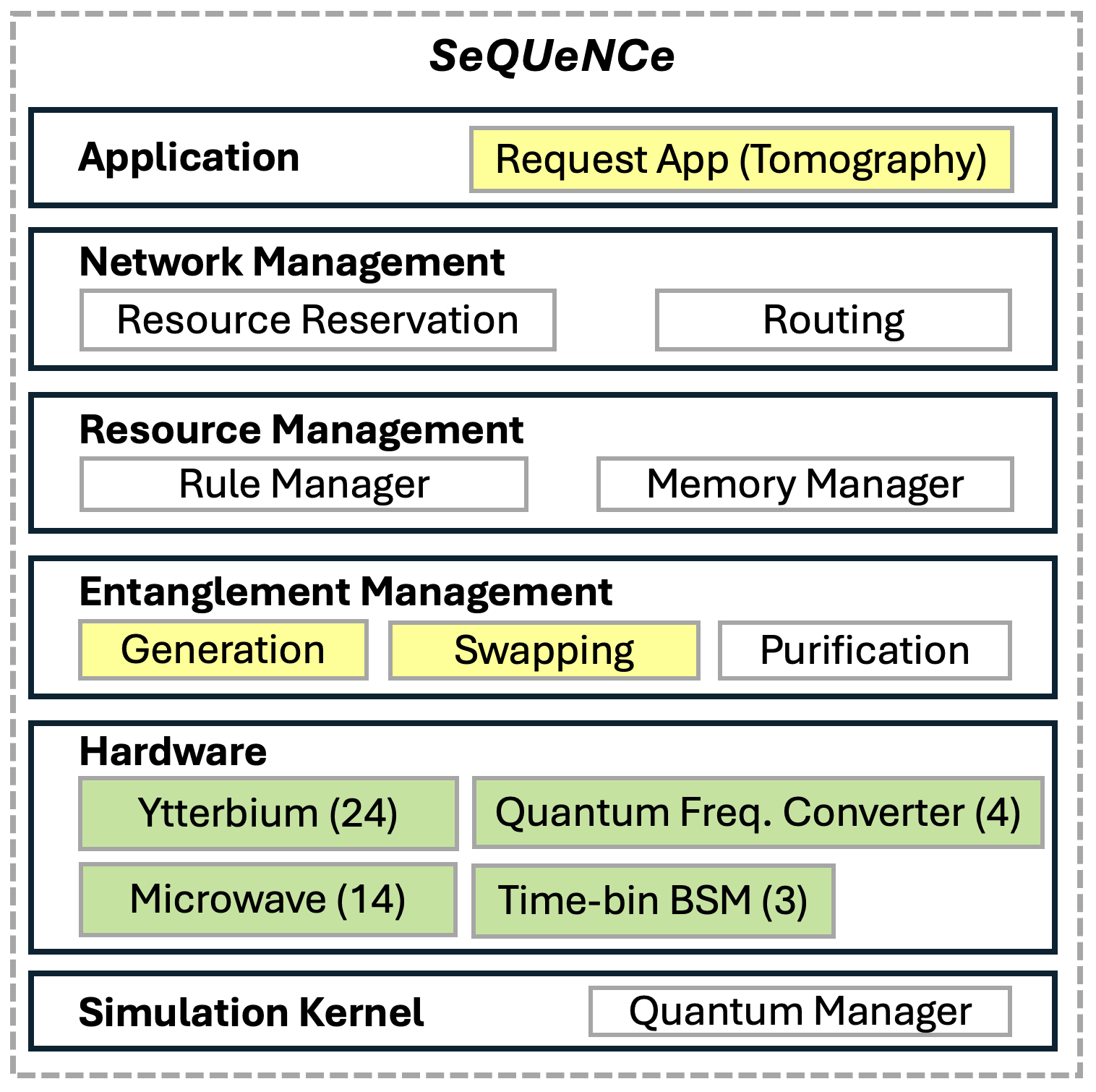}
    \vspace{-0.1in}
    \caption{SeQUeNCe has six modules, and each module has several components (not all components are shown). We introduced four new hardware models (green) and customized three components (yellow). The numbers in parenthesis indicate the amount of parameters for each new hardware module.}
    \vspace{-0.1in}
    \label{fig:sequence}
\end{figure}

In this section, we discuss our implementation of the new device models introduced in \S\ref{sec:device} in SeQUeNCe\cite{sequence}, showing how we customize SeQUeNCe to support heterogeneous quantum links and partial quantum state tomography.


\subsection{Hardware Module}

Four new classes are introduced for the four new hardware models (see green boxes in Figure~\ref{fig:sequence}). The new hardware classes are easy to integrate into SeQUeNCe's simulation architecture and interact with other modules such as Entanglement Generation.

\subsubsection{\yb} Class \yb\ encapsulates the 24 parameters of the Ytterbium memory and is inherited from SeQUeNCe's built-in \texttt{Memory} class. 
Each of the 24 physical parameters is an attribute of the \yb\ class. Among the 24 parameters, we evaluate the following three in \S\ref{sec:result}:
(1) \texttt{photon\_collection\_efficiency} (the percentage of photons that successfully enter the optical fiber after being emitted), 
(2) \texttt{attempts\_per\_reload} (the number of attempts before reloading the Ytterbium atoms), and
(3) \texttt{time\_bin\_width} (the width of the time-bin).

Apart from those 24 attributes, there is a \texttt{qstate\_key} attribute, a key for SeQUeNCe's built-in \texttt{QuantumManager} class that points to a state vector describing the state of the Ytterbium qubit. 
Besides the attributes, the \yb\ class has the following methods to perform the \yb-photon generation steps:




\begin{itemize}
    \item \texttt{initialize\_cool\_prep()} handles four steps: Reset, Initialization, Cooling, and Preparation.
    The Reset occurs each \texttt{attemps\_per\_reload} number of attempts, reloading the atoms and taking \texttt{reload\_time} ($500~ms$).

    During this operation, if the \texttt{atom\_state} attribute is $lost$, \texttt{atom\_state} will be reset to $\ket{e}$ due to reloading. Additionally, the class's attribute \texttt{efficiency} is reset to its default value (it had been set to $0$ while atom was lost). 
    For initialization, this method requires \texttt{initialize\_time} ($51.4~\mu s)$, and there is a $3\%$ probability that \texttt{atom\_state} is set to $lost$. 
    Cooling is an additional requirement of \texttt{cooling\_time} ($1.4~ms$). Additionally, the preparation portion of this method takes \texttt{state\_prep\_time} ($5.3~\mu s$) to set the state vector pointed to by the \texttt{Yb}'s \texttt{qstate\_key} to $\frac{1}{\sqrt 2}(\ket{0} + \ket{1})$, which corresponds to $\frac{1}{\sqrt 2}(\ket{e_\uparrow} + \ket{e_\downarrow})$.
    
    \item \texttt{excite()} comes from parent class \texttt{Memory} and handles the Generation step. This method first calls another class method \texttt{atom\_transition()}, and based on the branching ratios for each of the decay channels from $^3D_1$, it sets \texttt{atom\_state} to $\ket{e},\ket{g},$ or $lost$. 
    If the atom is lost, \texttt{efficiency} is set to $0$. \texttt{atom\_transition()} returns the wavelength of the underwent decay channel. 
    In \texttt{excite()}, one of SeQUeNCe's built-in \texttt{Photon} objects is created with the returned wavelength. This \texttt{Photon} is attached with loss from the \texttt{photon\_collection\_efficiency} and from a function of our parameters: 
    \texttt{late\_decay\_prob} = $exp(-\frac{\texttt{bin\_width}}{\texttt{state\_lifetime}})$. This late decay probability is the probability that the atom does not decay from $^3D_1$ (\texttt{state\_lifetime} is its lifetime) within the time-bin (\texttt{bin\_width}). 
    Additionally, the \texttt{Photon} object contains the same \texttt{qstate\_key} as its parent memory, indicating atom-photon entanglement. 
    The \texttt{Photon} object is then transmitted into one of SeQUeNCe's built-in \texttt{QuantumChannel} objects, set with an attenuation of $0.3\frac{dB}{km}$ \cite{covey}. 
    As it is impossible to simulate superposition of time in a discrete-event simulator, 
    \blue{we choose to schedule this transmission event at the early time-bin. This allows early and late detection events to be handled within the \texttt{TimeBinBSM} module.}

\end{itemize}

\begin{figure*}
    \centering
    \includegraphics[width=\linewidth]{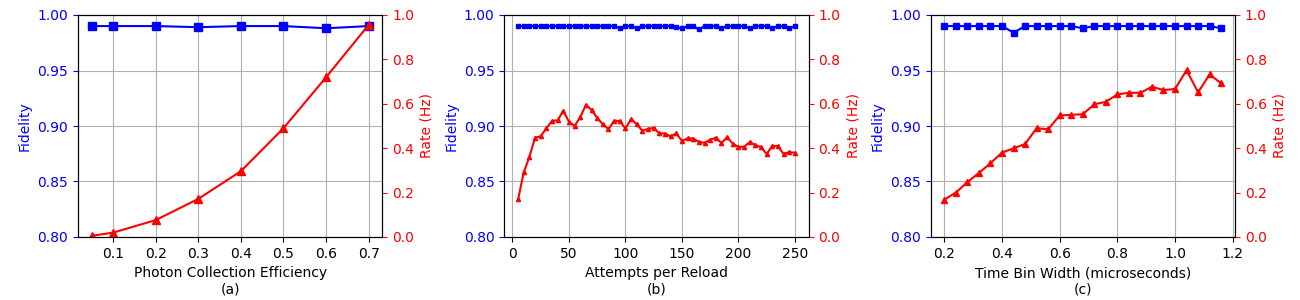}
    \vspace{-0.3in}
    \caption{Simulation results for a \yb-\yb\ link. (a) varies the photon collection efficiency; (b) varies the attempts per reload; (c) varies the time-bin width. }
    \vspace{-0.1in}
    \label{fig:yb-yb}
\end{figure*}

\begin{figure*}
    \centering
    \includegraphics[width=\linewidth]{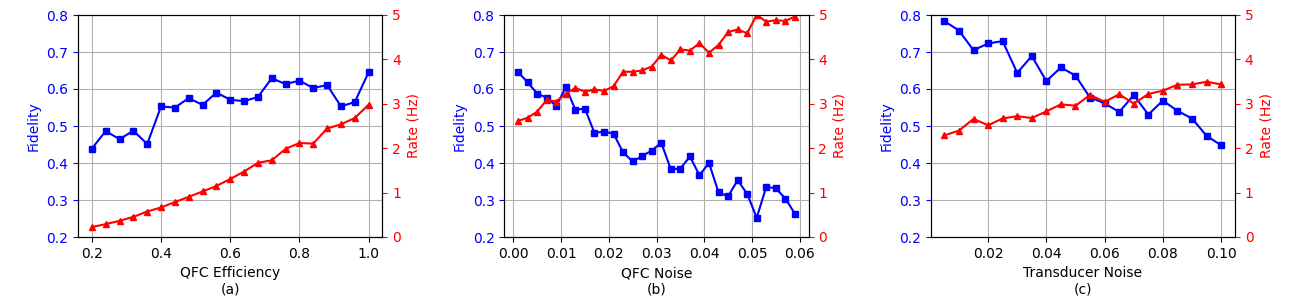}
    \vspace{-0.3in}
    \caption{Simulation results for a \yb-\uw\ link. (a) varies the QFC efficiency; (b) varies the QFC noise; (c) varies the transducer noise.}
    \vspace{-0.1in}
    \label{fig:yb-uw}
\end{figure*}

\subsubsection{\uw} Class \uw\ encapsulates the 14 parameters of the microwave memory and is inherited from SeQUeNCe's \texttt{Memory} built-in class. 
Each of the 14 physical parameters is an attribute of the \uw\ class. 
Among the 14 parameters, we evaluated the following two in \S\ref{sec:result}:
(1) \texttt{transducer\_noise} (the average number of noise photons emitted each transduction attempt) and
(2) \texttt{coherence\_time} (the coherence time of the transmon).
Besides the attributes, the \uw\ class has the following methods to perform the \uw-photon generation:
\begin{itemize}
    \item \texttt{initialize\_prep()} updates the state vector pointed to by the \uw's \texttt{qstate\_key} to $\frac{1}{\sqrt2}(\ket{0} + \ket{1})$ taking time of \texttt{initialize\_time} ($500~\mu s$) and \texttt{state\_prep\_time} ($30~ns$).
    \item \texttt{excite()}: one of SeQUeNCe's built-in \texttt{Photon} objects is created with the aforementioned wavelength. This \texttt{Photon} is attached with loss and noise from the \texttt{transduce()} method and later sent to the fiber at the early time-bin. 
    \item \texttt{transduce()} takes in a photon and adds loss (1 - \texttt{transducer\_efficiency}) and noise (a \blue{geometric sampling around the mean of \texttt{transducer\_noise})}.
    
\end{itemize}

\subsubsection{\texttt{QFC}} Class \texttt{QFC} encapsulates the four parameters of the quantum frequency converter and directly inherits from SeQUeNCe's built-in \texttt{Entity} class, the fundamental unit of the simulation world. The two most significant parameters being evaluated in \S\ref{sec:result} are (1) \texttt{qfc\_efficiency} (the probability of successfully converting a photon) and 
(2) \texttt{qfc\_noise} (the probability of generating a noise photon \blue{across} the early and late time-bins).

\subsubsection{\texttt{TimeBinBSM}} Class \texttt{TimeBinBSM} encapsulates the three parameters of the Time-bin BSM node and is inherited from SeQUeNCe's built-in \texttt{BSM} class. \blue{
It measures the state vector of each incoming \texttt{Photon}, records the time of a detector `click', and sends the `click' timestamp to end nodes.
}

\subsection{Entanglement Management Module}
For entanglement generation, we customize SeQUeNCe's built-in class \texttt{EntanglementGenerationA} by overriding the following methods:
\begin{itemize}
    \item \texttt{update\_memory()}: checks if there is exactly one click in each time-bin and signals entanglement succeeded if so, and failed otherwise. If the source of both clicks are signal photons, both memory's \texttt{qstate\_key}'s are pushed to point to $\psi^{+}$. 
    If one or more clicks are from noise photons, we have false positive entanglement, and the qubit states are not entangled.
    \item \texttt{emit\_event()}: calls \texttt{initialize\_cool\_prep()} and \texttt{excite()} on the relevant memory object to ensure possible photon emission at the start of the early time-bin.
    \item \texttt{received\_message()}: generates agreement over emit time, bin separation, and bin width of two heterogeneous end nodes with different generation cycles.
\end{itemize}

To generate entanglement between remote nodes, entanglement swapping is necessary, thus we reused SeQUeNCe's built-in class \texttt{EntanglementSwappingA} and \texttt{EntanglementSwappingB} with minimal changes.

\subsection{Application Module}
After successful entanglement generation, we estimate a lower bound on the entangled pair's fidelity to \blue{${\psi^{+}}$} via partial quantum state tomography on a customized \texttt{RequestApp} in the application module. 
This is necessary because the \texttt{Yb} device cannot perform $Y$-basis measurements.
Thus, we generate a lower bound on entanglement fidelity by alternating between $X$ and $Z$ basis measurement and using the formula:
$$\mathcal{F}  \geq \mathcal{M}*\frac{1}{2}(\rho^Z_{22} + \rho^Z_{33} + \rho^X_{11} + \rho^X_{44} - \rho^X_{22} - \rho^X_{33} - 2\sqrt{\rho^Z_{11}\rho^Z_{44}})$$
from \cite{fidelity}. Here, $\mathcal{M}$ is the product of the measurement fidelities of each of the entangled devices. \blue{Each $\rho^{Z,X}$ is an element of our two-qubit density matrix represented in the $Z$ or $X$ basis respectively.}

\section{Simulation Result}
\label{sec:result}

In this section, we present our simulation results for three scenarios: \yb-\yb\ link entanglement, heterogeneous \yb-\uw\ link entanglement, and heterogeneous \uw-\yb-\uw\ remote entanglement. Table~\ref{tab:sim-parameter} summarizes the key parameters and simulation settings. 
Timing parameters are omitted from the table and can be found in \S\ref{sec:device}. 
When one variable is swept in an experiment, all other parameters are set to the (default) value in Table~\ref{tab:sim-parameter}. \blue{Notably, some of these parameters assume expected future hardware improvements.}
Communication through the fiber contributes a non-negligible latency; we model the communication latency as:
\begin{equation}
    l = \frac{d}{c} + D_\mathrm{processing},
    \label{eqn:latency}
\end{equation}
where $d$ is the distance between two nodes, $c$ is the speed of light in optical fiber ($2\times10^8$m/s), and $D_\mathrm{processing}$ is the processing delay (10~$\mu s$).

\begin{table}[ht]
    \vspace{-0.1in}
    \caption{Selected Simulation (Default) Parameters}
    \vspace{-0.1in}
    \centering
    \begin{tabular}{|p{0.31\columnwidth}|c||p{0.3\columnwidth}|c|}
        \hline
         Photon Collection Eff. & 50\% & Attempts per Reload & 128   \\
         \hline
         Time-bin Width & 520~$ns$ & Detector Efficiency & 85\%  \\
         \hline
         Transducer Efficiency & 60\% &  Transducer Noise & 4.7\%\\
         \hline
         QFC Efficiency & 99\% & QFC noise & 0.5\% \\
         \hline
          Link distance & $1km$  & \uw\ Coherence Time & 500$\mu s$ \\
         \hline
    \end{tabular}
    \vspace{-0.1in}
    \label{tab:sim-parameter}
\end{table}

\subsection{\yb-\yb\ Link}

For the \yb-\yb\ link (see Fig.~\ref{fig:yb-yb}), we varied three significant parameters: photon collection efficiency, attempts per reload, and the time-bin width. In general, we observe the fidelity is almost invariant ($\sim99\%$ on average), because the only sources of infidelity are detector dark counts and the \yb\ atom falling out of the trap after photon emission. Such events have small probability and thus minimal impact on fidelity.

Regarding entanglement rate, Fig.~\ref{fig:yb-yb}(a) shows an expected rapid growth as we increase the photon collection efficiency (PCE), while
Fig.~\ref{fig:yb-yb}(b) shows that $\sim65$ attempts per reload provides an optimal entanglement rate of $0.6~Hz$. 
This curve first increases rapidly, and then slowly decreases. For low attempts per reload ($<30$), the device conducts time-costly reload operations while the \yb\ atom is very likely still in the trap, significantly harming rate. On the other hand, for high attempts per reload ($\gg 65$) the device is repeatedly attempting to entangle atoms that are very likely no longer in the trap, wasting time on cooling each attempt. 
The sharper fall-off on the left hand side of the plot (compared to the right hand side) is the result of reload time ($500~ms$) being much greater than cooling time ($1.4~ms$).
In Fig.~\ref{fig:yb-yb}(c), we found that increasing time-bin width from $0.2-1.2~ms$ increases rate, as a larger width increases the probability that the \yb\ emits a photon within the time-bin. However, the increased rate coming from a greater bin width is matched with an increased susceptibility to noise in the time-bin (not shown in Fig.~\ref{fig:yb-yb}(c)), which is of consideration in schemes with higher noise levels.

\subsection{\yb-\uw\ Link}

For the \yb-\uw\ link we vary QFC efficiency, QFC noise, and transducer noise due to their significant impact on entanglement rate and fidelity.
In Fig.~\ref{fig:yb-uw}(a), we observe that an increase in QFC efficiency ($0.2 \rightarrow 1.0$) leads to an increase in both fidelity ($0.45 \rightarrow 0.65$) and rate ($0.2 \rightarrow 3 ~Hz$). 
The increase in rate is straightforward, while the increase in fidelity is because QFC noise, which is not subject to QFC efficiency, is being held constant while the efficiency is being changed. Low efficiency means a higher proportion of clicks are from noise and vice versa. This higher number of noise clicks leads to false positive entanglement heralding and decreased fidelity.
In both Figures~\ref{fig:yb-uw}(b) and (c), we see an increase in QFC/transducer noise causing a decrease in fidelity and an increase in rate. The decrease in fidelity is straightforward, and the increase in rate is paired with that, as more noise leads to more false positives. 

\subsection{\uw-\yb-\uw\ Remote Entanglement}

\begin{figure}
    \centering
    \includegraphics[width=\linewidth]{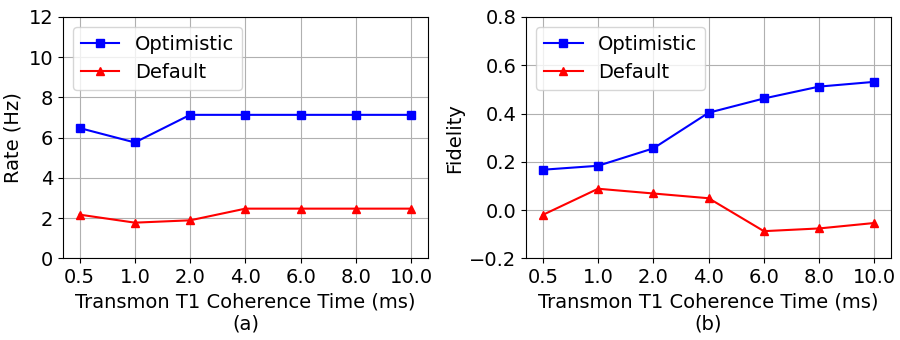}
    \vspace{-0.3in}
    \caption{Simulation results for the \uw-\yb-\uw\ remote entanglement generation. (a) and (b) both vary the coherence time of the \uw\ node and compare the results for default parameters and optimistic parameters.}
    \vspace{-0.2in}
    \label{fig:uw-yb-uw}
\end{figure}

In  Figure~\ref{fig:uw-yb-uw}, we study the entanglement generation fidelity and rate between two \uw\ nodes via a \yb-based quantum repeater as we vary the transmon's $T_1$ coherence time from its default of $0.5~ms$ to a maximum of $10~ms$. We consider two different parameter regimes: \blue{optimistic}
(all efficiency parameters are $100\%$ and noise rates are $0\%$) and default (default model parameters). 
We note that in the \blue{optimistic} case, there is still loss due to unavoidable transmon decoherence, the \yb\ branching ratios, \yb\  decay outside of the time-bin, and optical channel attenuation. 
During entanglement swapping, one link will herald entanglement first, and before swapping and tomography can occur, it must wait for the other link to also become entangled. This can take far longer than the coherence time of the already entangled transmon, leading to possible decoherence and a non-entangled final pair. 

Figure~\ref{fig:uw-yb-uw}(a) details the effect transmon coherence time has on rate. As expected, the rate is much higher for the \blue{optimistic} case ($\sim 7~Hz$) than for the default parameters ($\sim 2~Hz$). The rate is largely constant in both cases because the decoherence only impacts rate when it occurs during the generation step, which is very improbable. Decoherence after generation will not impact whether our click pattern suggests we have entanglement, but it will hurt fidelity.
In Figure~\ref{fig:uw-yb-uw}(b), we see the impact of transmon coherence on fidelity. For default parameters, we notice fidelity just oscillates around $0$ as irrespective of coherence time, our rate is low enough that the earliest of our transmons to become entangled nearly always decoheres before the other link generates entanglement. This shows that significant hardware improvements are needed to achieve usable fidelities. The negative numbers are a function of our fidelity formula being a lower bound, not fidelity itself.

\blue{We note that the difference between true fidelity and our lower bound stems from the $\sqrt{\rho^Z_{11}\rho^Z_{44}}$ term \cite{fidelity}. Thus, when those matrix elements are prevalent, our fidelity lower bound is worse (i.e., for a completely mixed state, true fidelity is $\mathcal{F} = 0.25$ while our lower bound computes $\mathcal{F} = 0$). 
The inaccuracy of our bound in the low-fidelity region is acceptable, as states in this regime are unusable no matter what.} In the \blue{optimistic} case, due to the higher rate, increased coherence time leads to tangible increases in fidelity (up to $>0.5$ at $T_1 =10~ms$) due to less decoherence during the time the first-entangled transmon has to wait for the other link to entangle.


\section{Conclusion}
\label{sec:conclusion}

In this paper, we used SeQUeNCe to develop faithful physical device models for a \yb\ neutral atom node, a \uw\ node composed of a transmon and a quantum transducer, a quantum frequency converter, and a Bell state measurement device. We also developed a heterogeneous entanglement generation protocol and simulated \yb-\yb\ and \yb-\uw\ links as well as a heterogeneous \yb-\uw-\yb\ network. \blue{Most significantly, we observed that entanglement fidelity across edge-nodes in a linear network was highly correlated with the shortest coherence time of a node. This is expected, as two sets of entangled qubits are required for swapping, and when the entanglement rate is low, the time between the first entanglement and the second can exceed the coherence time of the qubits -- thus degrading our entanglement fidelity.}

\blue{As this work is open source, and the heterogeneous entanglement generation protocol can stack with any hardware model inheriting from SeQUeNCe's \texttt{HetQR} (Heterogeneous Quantum Router) module, we hope this software enables developers to model larger quantum networks composed of new devices and topologies.}




\bibliographystyle{IEEEtran}
\bibliography{bib}

@IEEEtranBSTCTL{IEEEexample:BSTcontrol,
CTLuse_forced_etal       = "yes",
CTLmax_names_forced_etal = "1",
CTLnames_show_etal       = "1" }

@article{wehner2018quantum,
  title={Quantum internet: A vision for the road ahead},
  author={Wehner, Stephanie and Elkouss, David and Hanson, Ronald},
  journal={Science},
  volume={362},
  year={2018},
  publisher={American Association for the Advancement of Science}
}

@article{hu-teleportation-nrp23,
    author = {Hu, Xiaomin and Guo, Yu and Liu, Biheng and Li, Chuanfeng and Guo, Guangcan},
    title = {Progress in quantum teleportation},
    journal = {Nature Reviews Physics},
    year = {2023},
    volume = {5},
    doi = {https://doi.org/10.1038/s42254-023-00588-x}
}

@article{calaffi-dqc-cn24,
title = {Distributed quantum computing: A survey},
journal = {Computer Networks},
volume = {254},
year = {2024},
doi = {https://doi.org/10.1016/j.comnet.2024.110672},
author = {Marcello Caleffi and Michele Amoretti and Davide Ferrari and Jessica Illiano and Antonio Manzalini and Angela Sara Cacciapuoti},
}

@article{hillery-qsn-pra23,
title = {Discrete outcome quantum sensor networks},
journal = {Phys. Rev. A},
volume = {107},
publisher = {American Physical Society},
author = {Hillery, Mark and Gupta, Himanshu and Zhan, Caitao},
year = {2023},
doi = {10.1103/PhysRevA.107.012435}
}

@article{bennett-qkd-14,
   title={Quantum cryptography: Public key distribution and coin tossing},
   volume={560},
   journal={Theoretical Computer Science},
   author={Bennett, Charles H. and Brassard, Gilles},
   year={2014}
}

@inproceedings{quantnet-24,
    title = {The {QUANT}-{NET} {Testbed} {Development} and {Preliminary} {Results}},
    booktitle = {IEEE QCE},
    author = {Schon, Damian and Umesh, Prathwiraj and Cheah, You-Wei and Yu, Se-Young and Kissel, Ezra and Valivarthi, Raju and Saglamyurek, Erhan and Ramakrishnan, Lavanya and Wu, Wenji and Sipahigil, Alp and Spiropulu, Maria and Häffner, Hartmut and Monga, Inder},
    year = {2024},
}

@article{oakridge-prx-21,
  title = {Reconfigurable Quantum Local Area Network Over Deployed Fiber},
  author = {Alshowkan, Muneer and Williams, Brian P. and Evans, Philip G. and Rao, Nageswara S.V. and Simmerman, Emma M. and Lu, Hsuan-Hao and others},
  journal = {PRX Quantum},
  volume = {2},
  issue = {4},
  year = {2021},
}

@misc{longisland-24,
      title={A long-distance quantum-capable internet testbed}, 
      author={Dounan Du and Leonardo Castillo-Veneros and Dillion Cottrill and Guo-Dong Cui and Paul Stankus and Dimitrios Katramatos and Julián Martínez-Rincón and Eden Figueroa},
      year={2024},
}

@misc{interqnet-25,
      title={{InterQnet}: A Heterogeneous Full-Stack Approach to Co-designing Scalable Quantum Networks}, 
      author={Joaquin Chung and Daniel Dilley and Ely Eastman and Alvin Gonzales and Kara Hokenstad and Md Shariful Islam and Varun Jorapur and Joseph Petrullo and others},
      year={2025},
      eprint={2509.19503},
      archivePrefix={arXiv},
      primaryClass={quant-ph},
}

@article{sequence,
author = {Xiaoliang Wu and Alexander Kolar and Joaquin Chung and Dong Jin and Tian Zhong and Rajkumar Kettimuthu and Martin Suchara},
title = {{SeQUeNCe}: a customizable discrete-event simulator of quantum networks},
journal = {Quantum Science and Technology},
volume = {6},
year = {2021},
doi = {10.1088/2058-9565/ac22f6},
publisher = {IOP Publishing},
}

@inproceedings{quisp,
   title={{QuISP}: a Quantum Internet Simulation Package},
   DOI={10.1109/qce53715.2022.00056},
   booktitle={IEEE QCE},
   author={Satoh, Ryosuke and Hajdusek, Michal and Benchasattabuse, Naphan and Nagayama, Shota and Teramoto, Kentaro and Matsuo, Takaaki and others},
   year={2022},
}

@article{netsquid,
   title={{NetSquid}, a NETwork Simulator for QUantum Information using Discrete events},
   DOI={10.1038/s42005-021-00647-8},
   journal={Communications Physics},
   author={Coopmans, Tim and others},
   year={2021},
}

@INPROCEEDINGS{acp-qcnc-25,
  author={Zhan, Caitao and Chung, Joaquin and Zang, Allen and Kolar, Alexander and Kettimuthu, Rajkumar},
  booktitle={QCNC}, 
  title={Design and Simulation of the Adaptive Continuous Entanglement Generation Protocol}, 
  year={2025},
}

@article{swappingtree-tqe-22,
   title={Efficient Quantum Network Communication Using Optimized Entanglement Swapping Trees},
   journal={IEEE TQE},
   author={Ghaderibaneh, Mohammad and Zhan, Caitao and Gupta, Himanshu and Ramakrishnan, C. R.},
   year={2022}
}

@ARTICLE{graphstate-tqe-25,
  author={Fan, Xiaojie and Zhan, Caitao and Gupta, Himanshu and Ramakrishnan, C. R.},
  journal={IEEE Transactions on Quantum Engineering}, 
  title={Optimized Distribution of Entanglement Graph States in Quantum Networks}, 
  year={2025},
}

@inproceedings{hetero-qce-25,
   title={Servicing Demands Between Multiple Pairs of Nodes on Heterogeneous Quantum Networks},
   booktitle={IEEE QCE},
   author={Anoop Kumar Pandey and Bheemarjuna Reddy Tamma and M V Panduranga Rao},
   year={2025},
}

@inproceedings{quantum-conext-20,
author = {Kozlowski, Wojciech and Dahlberg, Axel and Wehner, Stephanie},
title = {Designing a quantum network protocol},
year = {2020},
booktitle = {ACM CoNEXT},
}

@inproceedings{qlan-icc-25,
      title={{Simulation of Entanglement-Enabled Connectivity in QLANs using SeQUeNCe}}, 
      author={Francesco Mazza and Caitao Zhan and Joaquin Chung and Rajkumar Kettimuthu and Marcello Caleffi and Angela Sara Cacciapuoti},
      year={2025},
      booktitle = {IEEE ICC}
}

@inproceedings{transducer-qce-25,
      title={Simulation of Quantum Transduction Strategies for Quantum Networks}, 
      author={Laura d'Avossa and Caitao Zhan and Joaquin Chung and Rajkumar Kettimuthu and Angela Sara Cacciapuoti and Marcello Caleffi},
      year={2025},
      booktitle = {IEEE QCE}
}

@INPROCEEDINGS{atom-qce-23,
  author={Zhou, Ruilin and Lai, Xuanying and Gan, Yuhang and Obraczka, Katia and Du, Shengwang and Qian, Chen},
  booktitle={IEEE QCE}, 
  title={A Simulator of Atom-Atom Entanglement with Atomic Ensembles and Quantum Optics}, 
  year={2023},
}

@INPROCEEDINGS{absorptive-qce-22,
  author={Zang, Allen and Kolar, Alexander and Chung, Joaquin and Suchara, Martin and Zhong, Tian and Kettimuthu, Rajkumar},
  booktitle={2022 QCE}, 
  title={Simulation of Entanglement Generation between Absorptive Quantum Memories}, 
  year={2022},
}

@article{covey,
    author = {Li, Lintao and Hu, Xiye and Jia, Zhubing and Huie, William and Sun, Won Kyu Calvin and Aakash and Dong, Yuhao and Hiri-O-Tuppa, Narisak and Covey, Jacob P.},
    title ={Parallelized telecom quantum networking with a ytterbium-171 atom array} ,
    journal = {arXiv preprint arXiv:2502.17406},
    year = {2025}
}

@article{neutralatom,
  title = {Quantum computing with neutral atoms},
  author = {Henriet, Lo{\"{i}}c and Beguin, Lucas and others},
  journal = {{Quantum}},
  publisher = {{Verein zur F{\"{o}}rderung des Open Access Publizierens in den Quantenwissenschaften}},
  volume = {4},
  year = {2020}
}

@article{transmon,
  title = {Quantum Communication with Time-Bin Encoded Microwave Photons},
  author = {Kurpiers, P. and Pechal, M. and Royer, B. and Magnard, P. and Walter, T. and others},
  journal = {Phys. Rev. Appl.},
  volume = {12},
  issue = {4},
  year = {2019},
  month = {Oct},
  publisher = {American Physical Society},
}

@article{transducer-review,
author = {Han, Xu and Fu, Wei and Zou, Chang-Ling and Jiang, Liang},
year = {2021},
month = {08},
pages = {1050-1064},
title = {Microwave-optical quantum frequency conversion},
volume = {8},
journal = {Optica},
doi = {10.1364/OPTICA.425414}
}

@article{bersin_development_2024,
    title = {Development of a {Boston}-area 50-km fiber quantum network testbed},
    volume = {21},
    journal = {Physical Review Applied},
    author = {Bersin, Eric and Grein, Matthew and Sutula, Madison and Murphy, Ryan and Huan, Yan Qi and Stevens, Mark and Suleymanzade, Aziza and Lee, Catherine and Riedinger, Ralf and Starling, David J. and Stas, Pieter-Jan and Knaut, Can M. and Sinclair, Neil and Assumpcao, Daniel R. and Wei, Yan-Cheng and Knall, Erik N. and Machielse, Bartholomeus and Sukachev, Denis D. and Levonian, David S. and Bhaskar, Mihir K. and Lončar, Marko and Hamilton, Scott and Lukin, Mikhail and Englund, Dirk and Dixon, P. Benjamin},
    year = {2024},
}

@misc{heterogeneous_routing-25,
      title={Routing in Quantum Repeater Networks with Mixed Efficiency Figures}, 
      author={Vinay Kumar and Claudio Cicconetti and Marco Conti and Andrea Passarella},
      year={2025},
      eprint={2310.08990},
      archivePrefix={arXiv},
      primaryClass={quant-ph},
}

@article{uw_readout,
   title={Rapid High-Fidelity Single-Shot Dispersive Readout of Superconducting Qubits},
   volume={7},
   journal={Physical Review Applied},
   publisher={American Physical Society (APS)},
   author={Walter, T. and Kurpiers, P. and Gasparinetti, S. and Magnard, P. and Potočnik, A. and Salathé, Y. and Pechal, M. and Mondal, M. and Oppliger, M. and Eichler, C. and Wallraff, A.},
   year={2017},
}

@article{fidelity,
   title={Heralded entanglement between solid-state qubits separated by three metres},
   volume={497},
   ISSN={1476-4687},
   number={7447},
   journal={Nature},
   publisher={Springer Science and Business Media LLC},
   author={Bernien, H. and Hensen, B. and Pfaff, W. and Koolstra, G. and Blok, M. S. and Robledo, L. and Taminiau, T. H. and Markham, M. and Twitchen, D. J. and Childress, L. and Hanson, R.},
   year={2013},
}

@article{delft_testbed-science-24,
    author = {Arian J. Stolk  and Kian L. van der Enden  and Marie-Christine Slater  and Ingmar te Raa-Derckx  and Pieter Botma  and Joris van Rantwijk  and others},
    title = {Metropolitan-scale heralded entanglement of solid-state qubits},
    journal = {Science Advances},
    volume = {10},
    year = {2024},
}

@misc{github,
  author = {Miller, Hayden},
  title = {Heterogenous Quantum Network Simulator},
  year = {2026},
  howpublished = {\url{https://github.com/haydenmllr1317/heterogenous}},
  note = {GitHub repository}
}

\end{document}